\documentstyle[epsf]{aipproc}

\def\bu{B^+}
\def\bd{B^0_d} 
\def\bs{B^0_s}
\def\bmix{B^0 \mbox{--} \overline{B^0}}
\def\bdmix{B_d^0 \mbox{--} \overline{B_d^0}}
\def\bsmix{B_s^0 \mbox{--} \overline{B_s^0}}

\def\sinsqth{\sin^2\theta_W^{eff}}
\def\Zbb{Z^0 \rightarrow b{\overline b}}
\def\Zcc{Z^0 \rightarrow c{\overline c}}
\def\Zff{Z^0 \rightarrow f{\overline f}}

\begin{document}

\begin{flushright}
{\small
SLAC--PUB--7719\\
December 1997\\}
\end{flushright}

\title{Electroweak and Heavy Flavor Physics at SLD}

\author{St\'ephane Willocq$^*$}
\address{$^*$Stanford Linear Accelerator Center\thanks{Work
supported in part  by DOE Contract DE-AC03-76SF00515(SLAC).}\\
Stanford University, Stanford, CA 94309 \\[0.4cm] ~~\\
Representing the SLD Collaboration \\
Stanford Linear Accelerator Center \\
Stanford University, Stanford, CA 94309}

\maketitle

\begin{abstract}
We review recent electroweak and B physics results obtained
in polarized $e^+ e^-$ interactions at the SLC
by the SLD experiment.
Unique and precise measurements of the electroweak parameters
$A_e$, $A_b$, $A_c$, $R_b$ and $R_c$
provide powerful constraints on the Standard Model.
The excellent 3-D vertexing capabilities of SLD are further exploited
to extract precise $\bu$ and $\bd$ lifetimes, as well as
measurements of the time evolution of $B^0 - \overline{B^0}$ mixing.
\end{abstract}

\begin{center}

\vspace{25mm}
{\sl Presented at the Workshop on Physics at the First Muon
Collider and at the Front End of a Muon Collider,
6-9 November 1997, Fermilab, Batavia, IL.}

\end{center}

\pagebreak

\title{Electroweak and Heavy Flavor Physics at SLD}

\author{St\'ephane Willocq}
\address{Stanford Linear Accelerator Center\thanks{Work
supported in part  by DOE Contract DE-AC03-76SF00515(SLAC).}\\
Stanford University, Stanford, CA 94309}


\maketitle

\begin{abstract}
We review recent electroweak and B physics results obtained
in polarized $e^+ e^-$ interactions at the SLC
by the SLD experiment.
Unique and precise measurements of the electroweak parameters
$A_e$, $A_b$, $A_c$, $R_b$ and $R_c$
provide powerful constraints on the Standard Model.
The excellent 3-D vertexing capabilities of SLD are further exploited
to extract precise $\bu$ and $\bd$ lifetimes, as well as
measurements of the time evolution of $B^0 - \overline{B^0}$ mixing.
\end{abstract}

\section*{Introduction}

The various measurements presented below rely on the strengths of
the SLC/SLD environment.
Most important is the fact that the electrons are longitudinally
polarized at the interaction point. Average polarizations of
$(63.0 \pm 1.1)\%$, $(77.2 \pm 0.5)\%$ and $(76.5 \pm 0.8)\%$
were measured during the 1993, 1994--95, and 1996 data taking periods
with a Compton Polarimeter~\cite{Woods}.
The numbers of hadronic $Z^0$ decays collected during these periods
are approximately 50K, 100K, and 50K, respectively.
A description of the detector can be found in Ref.~\cite{Rb}
and references therein.

\section*{Electroweak Physics}
\label{Sec_EW}

  In the Standard Model (SM), the tree-level differential cross section for
$e^+ e^- \rightarrow Z^0 \rightarrow f \bar{f}$
is expressed by
\begin{equation}
  \frac{d\sigma}{d\cos\theta} = \left(1-P_e A_e\right)
  \left(1 + \cos^2\theta\right) + 2\cos\theta\left(A_e - P_e\right) A_f ,
  \label{Equ_cross}
\end{equation}
where $\cos\theta$ is the cosine of the angle between the final state
fermion $f$ and the incident electron directions, $P_e$ is the electron
beam longitudinal polarization, and $A_e$ and $A_f$ are the asymmetry
parameters for the initial and final state fermions, respectively.
The parameter $A_f$ represents
the extent of parity violation at the $\Zff$ vertex and is defined as
$A_f = \frac{g^2_L - g^2_R}{g^2_L + g^2_R}$,
with the left- and right-handed coupling constants
$g_L$ and $g_R$.

The existence of parity violation introduces a forward-backward asymmetry
$A^{FB}_f = (\sigma_f^F - \sigma_f^B)/(\sigma_f^F + \sigma_f^B)$
which is equal to $\frac{3}{4} A_e A_f$.
At the SLC, the electron beam polarization allows
a left-right forward-backward asymmetry to be measured
\begin{equation}
  \tilde{A}^{FB}_f = \frac{\left[\sigma_f^F - \sigma_f^B\right]^{\mbox{left}} -
                     \left[\sigma_f^F - \sigma_f^B\right]^{\mbox{right}}}
                    {\left[\sigma_f^F + \sigma_f^B\right]^{\mbox{left}} +
                     \left[\sigma_f^F + \sigma_f^B\right]^{\mbox{right}}}
             = \frac{3}{4} P_e A_f.
\end{equation}
The latter asymmetry provides a {\em direct} measurement of $A_f$
and yields a statistical enhancement factor of $(P_e/A_e)^2 \simeq 25$
over the unpolarized forward-backward asymmetry.

A particularly powerful yet straightforward asymmetry is
the left-right cross-section
asymmetry $A_{LR}^0 = \frac{\sigma_L - \sigma_R}{\sigma_L + \sigma_R} = A_e$,
which yields a direct measurement of the coupling between the $Z^0$
and the $e^+ e^-$ initial state.

Measurements of the ratio between the partial $Z^0$ decay width
into $f\bar{f}$ and that into any hadron are also
sensitive to the $\Zff$ coupling constants
$R_f= \frac{\Gamma(Z\rightarrow f\bar{f})}{\Gamma(Z\rightarrow hadrons)}
\propto g^2_L + g^2_R$.

Precise measurements of $A_f$ and $R_f$ probe the effect
of radiative corrections to the $Z^0$ propagator
and the $Z f \bar{f}$ vertex.
Since the radiative corrections depend on the top and Higgs masses,
precise measurements can measure or constrain these quantities.
Furthermore, such measurements also are sensitive to physics beyond the SM.
Vacuum polarization corrections affect the value of
$\sin^2\theta_W^{eff}$ which is most precisely measured by the
left-right asymmetry
$A_{LR}^0 =
    \frac{2\left[1 - 4\sinsqth\right]}{1 + \left[1 - 4\sinsqth\right]^2}$.
Heavy quark partial widths and asymmetry parameters are most
sensitive to vertex corrections
but with different sensitivity to left- and right-handed coupling constants,
e.g., $R_b$ ($A_b$) is more sensitive to deviations
in the left- (right-) handed coupling.

  Electroweak measurements at SLD are only summarized here, for a complete
review of recent measurements see Ref.~\cite{Schumm}.

\subsection*{Left-Right Cross-Section Asymmetry}

  The measurement of $A^0_{LR}$ is a simple counting experiment.
One needs to count the number of $Z^0$ produced with left- and right-handed
electron beams, and measure the average electron beam polarization.
The event sample is selected using tracking and calorimetry information
and mostly consists of hadronic $Z^0$ decays (99.9\%).
In the 1996 data, the number of selected events is
28,713 and 22,662 for left- and right-handed electrons, respectively.
The resulting measured asymmetry is
$A_m = (N_L - N_R) / (N_L + N_R) = 0.1178 \pm 0.0044$(stat).
A very small correction $\delta = (0.08 \pm 0.08)\%$(syst)
is applied to take into account residual contamination in the
event sample and slight beam asymmetries.
Finally, the result is corrected for the electron beam polarization,
the initial and final state radiation
as well as for scaling the result to the $Z^0$ pole energy:
$A^0_{LR} = 0.1570 \pm 0.0057 \mbox{(stat)} \pm 0.0017 \mbox{(syst)}$,
giving
$\sin^2\theta_W^{eff} = 0.23025 \pm 0.00073 \mbox{(stat)}
       \pm 0.00021 \mbox{(syst)}$.
This preliminary measurement can be combined with previous measurements
from the 1992, 1993, and 1994--95 running periods \cite{ALR9495}
to yield $A^0_{LR} = 0.1550 \pm 0.0034$ and
$\sin^2\theta_W^{eff} = 0.23051 \pm 0.00043$.
This represents the most precise measurement of
$\sin^2\theta_W^{eff}$ by a single experiment and its uncertainty
is completely dominated by the statistical error.

\begin{figure}[htb]
  \centering
  \epsfxsize9cm
  \leavevmode
  \epsfbox{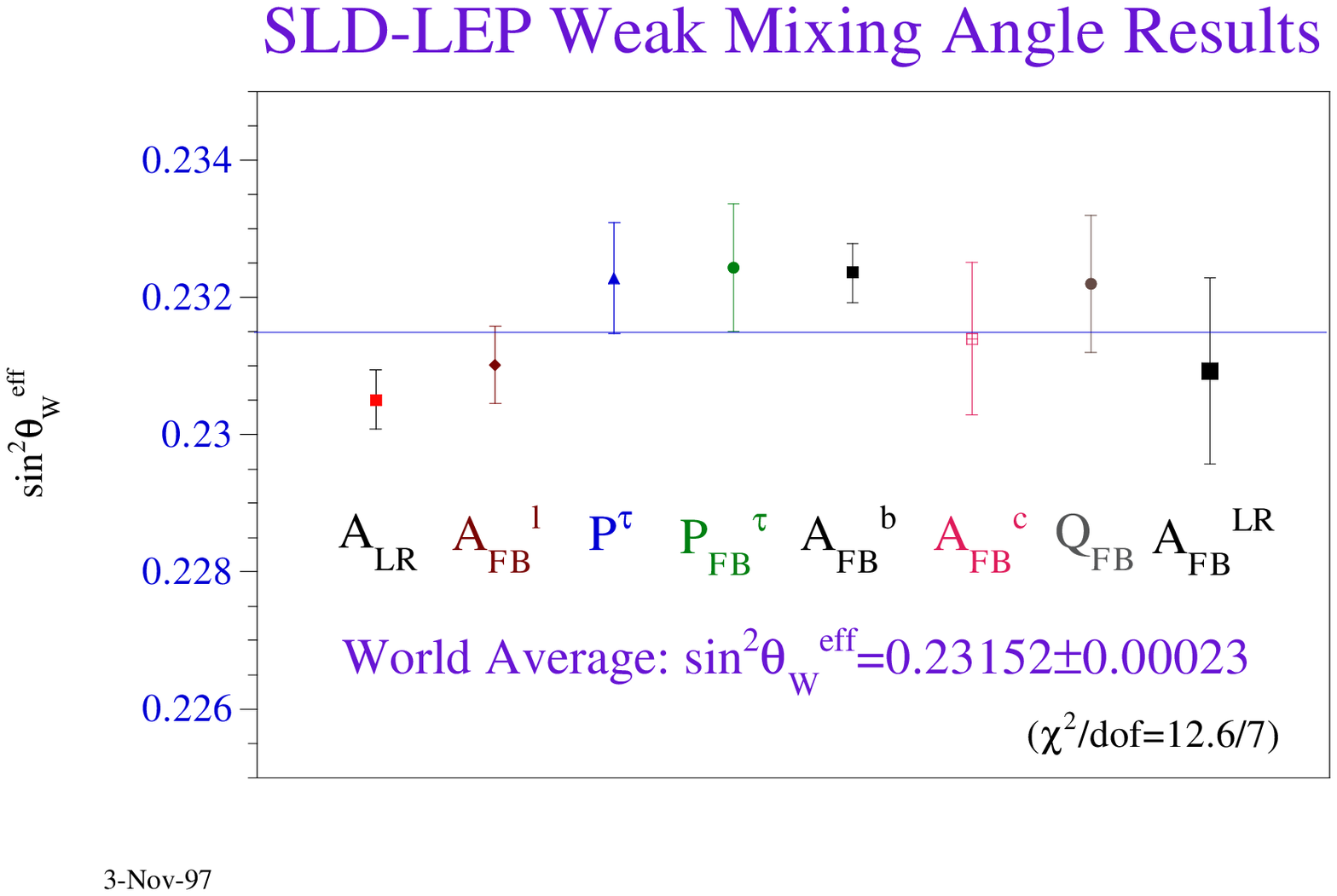}
  \caption{Measurements of $\sinsqth$ grouped by technique.}
  \label{Fig_sin2tw}
\end{figure}

Further information about $\sin^2\theta_W^{eff}$ can be obtained
from $Z^0$ decays into a pair of charged leptons ($\mu$ or $\tau$).
If lepton universality is assumed, the measurements may be combined to yield
\begin{eqnarray}
  \sin^2\theta_W^{eff} & = & 0.23055 \pm 0.00041.
\end{eqnarray}
This value can be compared with the
LEP average of $0.23162 \pm~0.00041$
obtained from lepton forward-backward asymmetries and
$\tau$ polarization measurements.
The SLD and LEP lepton averages agree within $1.9\:\sigma$.
Fig.~\ref{Fig_sin2tw} compares the SLD measurement with all others
obtained at LEP.

\subsection*{$Z$ Decay Partial Widths}
\label{Sec_Rbc}

  Another observable providing strong constraints on the SM is
$R_b = \Gamma(Z^0\rightarrow b\bar{b})/\Gamma(Z^0\rightarrow \mbox{hadrons})$.
The measurement proceeds by selecting hadronic $Z^0$ decays
and tagging each event hemisphere independently for the presence of
a $B$ hadron decay. By computing both the rate for tagging a hemisphere
and the rate for tagging both hemispheres in an event,
one can extract the value of $R_b$ and the hemisphere tag efficiency
$\epsilon_b$ from the data.
The Monte Carlo (MC) is used to estimate the charm and $uds$ efficiencies
as well as the small correlation between hemispheres.

  SLD has designed a new approach to achieve high-efficiency and
high-purity $b$ tagging. The excellent 3-D vertexing capabilities of the
vertex detector allow $B$ decays to be reconstructed with an inclusive
topological technique\cite{ZVTOP}.
This technique relies on the precise knowledge of the $e^+ e^-$
interaction point (IP), 7 $\mu$m (15-35 $\mu$m) perpendicular to (along)
the beam direction,
and the precise tracking achieved with the
pixel-based vertex detectors: VXD2 for 1993--95 and VXD3 for 1996 and
beyond.
The upgraded detector (VXD3) provides much improved polar angle coverage
($|\cos\theta|<0.90$), lever arm and self-tracking capabilities, as well
as a significantly reduced amount of material.
Secondary vertices are found in $50\%$ ($65\%$)
of $b$ hemispheres, $15\%$ ($20\%$)
of $c$ hemispheres but in less than 1\% of $uds$ hemispheres
for VXD2 (VXD3).
The $b$~hemisphere vertex finding efficiency increases with
the decay length $D$ to attain a constant level of 80\% for $D > 3$~mm.
Due to the typical $B \rightarrow D$ cascade structure of the decays,
not all tracks originate from a single space point.
Therefore, isolated tracks are attached to the vertex
if they extrapolate close to the vertex line-of-flight and
are sufficiently displaced from the IP.

\begin{figure}[htb]
  \centering
  \epsfxsize8cm
  \leavevmode
  \epsfbox{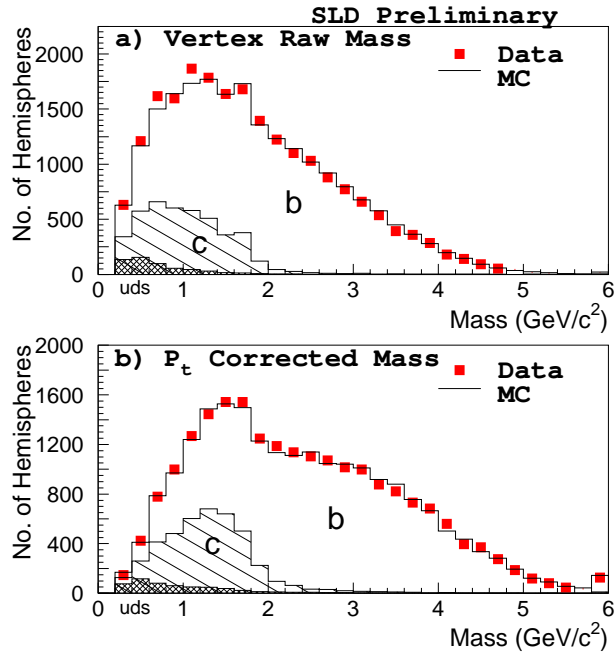}
  \caption{Distributions of (a) raw mass and (b) $p_T$-added mass
  for topological vertices in the 1993--95 data.}
  \label{Fig_bMass}
\end{figure}

  The mass of the reconstructed vertex is used to tag $b$ hemispheres.
A clear separation between $b$ and light-flavor hemispheres can be observed
in Fig.~\ref{Fig_bMass}(a).
In particular, the charm contribution vanishes above
the natural cutoff mass of $\sim$ 2 GeV.
Improved tagging efficiency
is achieved by constructing the $p_T$-added mass:
$M = \sqrt{M^2_{raw} + p^2_T} + |p_T|$,
where $p_T$ is the total momentum of all tracks in the vertex in the plane
perpendicular to the vertex axis. The value of $p_T$ is minimized
taking into account the uncertainties in the vertex and IP positions.
As a result, the $b$-tag efficiency is enhanced
by approximately 40\% without significant degradation in purity
[see Fig.~\ref{Fig_bMass}(b)].

Requiring $M > 2$ GeV yields
$b$-tag efficiencies of $(35.3 \pm 0.6)\%$ and $(47.9 \pm 0.8)\%$
for the 1993--95 and 1996 data, respectively. The corresponding samples
are 98\% pure in $B$ hadrons.
Combining all measurements yields
$R_b = 0.2124 \pm 0.0024 {\mbox{(stat)}} \pm 0.0017 {\mbox{(syst)}}$,
in agreement with the SM value of 0.2158.

  The above mass tag can be expanded to provide both a $b$ tag for
$M > 2$ GeV and a $c$ tag for $0.6 < M < 2.0$ GeV.
The $c$ tag is improved by utilizing the total momentum of tracks attached
to the vertex, exploiting the fact that charm hadrons in $c\bar{c}$ events
have a harder momentum spectrum than in $b\bar{b}$ events.
Similarly as was done for $R_b$, a system of five equations can be solved
for the various $b$- and $c$-tagging fractions.
Here, the values of $R_b$, $R_c$, as well as the $b$- and $c$-tag efficiencies
are extracted from the data.
The $c$-tag efficiencies are measured to be $(11.4 \pm 1.0)\%$ and
$(14.0 \pm 1.3)\%$ for 1993--95 and 1996 data, respectively,
with charm purities of 68\%.
Combining all data samples yields
$R_c = 0.181 \pm 0.012 {\mbox{(stat)}} \pm 0.008 {\mbox{(syst)}}$,
in agreement with the SM value of 0.172.
This double-tag technique has the advantage of having the lowest
systematic uncertainty of all current $R_c$ measurements.

\subsection*{Heavy Flavor Asymmetries}

  Different techniques have been used to measure the
asymmetry parameters $A_b$ and $A_c$.
These differ mostly by the method used to
determine which hemisphere contains the primary $b$ or $c$ quark:
jet charge, vertex charge, kaon charge, $D^{(\ast)}$ charge or lepton charge.
The first step (after hadronic event selection) in each analysis is
to tag $\Zbb$ ($\Zcc$) events either by applying the mass tag
utilized for the $R_b$ ($R_c$) analysis,
or by utilizing the fact
that leptons from $B$ hadron decays have harder $p$ and $p_T$
(with respect to the jet axis) distributions than those
from other processes.
$A_b$ and/or $A_c$ are determined with a fit to the $\cos\theta$-dependent
differential cross section [Eq.~(\ref{Equ_cross})].
The thrust axis direction is used to
provide the primary quark axis, except for the lepton analysis
which uses the jet axis.

  The first $A_b$ measurement tags $b$/$\bar{b}$ with
a momentum-weighted track charge defined as
$Q = -\sum Q_{track} \left|\vec{p} \cdot \hat{T}\right|^\kappa
  \mbox{sign}\left(\vec{p} \cdot \hat{T} \right)$,
where $\vec{p}$ is the three-momentum of each track and $Q_{track}$ its
charge, $\hat{T}$ is the direction of the event thrust axis,
signed such as to make $Q > 0$, and
the coefficient $\kappa = 0.5$.
The 100\% tag efficiency allows the analyzing power
of the tag to be calibrated directly from the data.
The probability to correctly tag
the primary $b$/$\bar{b}$ quark is
$P_{correct} = (1 + e^{-\alpha_b |Q|})^{-1}$,
where the constant $\alpha_b$ is measured to be $0.253 \pm 0.013$.
The average correct tag probability is thus
$\langle P_{correct} \rangle = 68\%$.

\begin{figure}[htb]
  \vspace{-14 mm}
  \centering
  \epsfxsize8.5cm
  \leavevmode
  \epsfbox{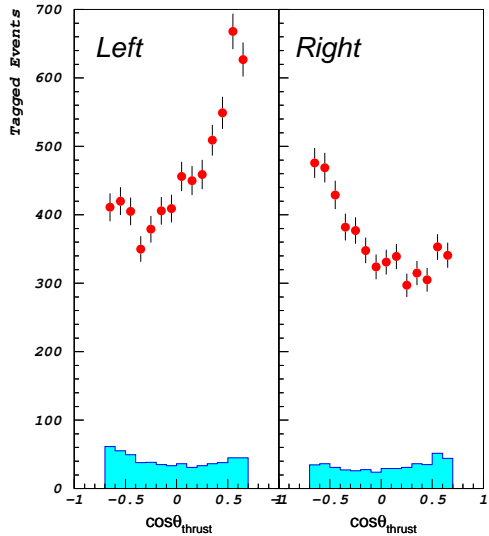}
  \caption{Distributions of the thrust axis $\cos\theta$ for
   left- and right-handed electrons.}
  \label{Fig_afbjetq}
\end{figure}

  Clear forward-backward asymmetries are then observed
for left- and right-handed electron beams, see Fig.~\ref{Fig_afbjetq}.

  The second $A_b$ measurement tags $b/\bar{b}$ by exploiting
the dominant $b \rightarrow c \rightarrow s$ transition
in $B$ decays.
Therefore, detection of $K^-$ ($K^+$) mesons tags $b$ ($\bar{b}$) quarks.
Right-sign kaon production has been measured by ARGUS\cite{ArgusK}
in $\bu$ and $\bd$ decays to be $(85 \pm 5)\%$ and $(82 \pm 5)\%$,
respectively.
The analysis presented here is the first application of kaon tagging
for a heavy quark asymmetry measurement.
Charged kaons are identified with the Cherenkov Ring Imaging
Detector and the rate of pion
misidentification is carefully calibrated from $K^0_s$ and $\tau$ decay
data samples.

  Another approach to heavy quark selection and flavor tagging
is to select leptons from semileptonic decays. The distinctive lepton total
and transverse momenta are exploited as well as the charge of the
lepton.

  The measured values of $A_b$ obtained with the three different tagging
techniques are presented in Table~\ref{Tbl_Ab}. The SLD average agrees
well with the SM prediction of 0.935.

\begin{table}
\caption{Measurements of the $\Zbb$ asymmetry parameter $A_b$.}
\label{Tbl_Ab}
\begin{tabular}{cccc}
 Jet Charge & Kaon Charge & Lepton Charge & SLD average \\
 \tableline
 $0.911 \pm 0.045 \pm 0.045$ & $0.877 \pm 0.068 \pm 0.047$ &
 $0.891 \pm 0.083 \pm 0.113$ & $0.898 \pm 0.052$ \\
 \hline
\end{tabular}
\end{table}

  We have also performed measurements of the asymmetry parameter $A_c$
in $\Zcc$ decays. The most powerful measurement relies on combined
kaon charge and vertex charge tags. This particular measurement based on
only 150K hadronic $Z^0$ decays is already among the best and should
become dominant with a modest increase in statistics.
$A_c$ is also measured with two more traditional techniques, as summarized
in Table~\ref{Tbl_Ac}. The SLD average agrees well with the SM prediction
of 0.67.

\begin{table}
\caption{Measurements of the $\Zcc$ asymmetry parameter $A_c$.}
\label{Tbl_Ac}
\begin{tabular}{cccc}
 Kaon \& Vtx Charge & Lepton Charge & $D^{(\ast)}$ Charge & SLD average \\
 \tableline
 $0.66 \pm 0.07 \pm 0.04$ & $0.61 \pm 0.10 \pm 0.07$ &
 $0.64 \pm 0.11 \pm 0.06$ & $0.647 \pm 0.060$ \\
 \hline
\end{tabular}
\end{table}

\section*{B Physics}
\label{Sec_bphys}

  Several aspects of the weak interaction can be probed by studying
the weak decays of $B$ hadrons. First, by measuring lifetimes, we
can test our understanding of $B$ hadron decay dynamics.
Second, we can
test the Cabibbo-Kobayashi-Maskawa (CKM) quark mixing matrix
description within the SM.

\subsection*{$\bu$ and $\bd$ Lifetimes}
\label{Sec_life}

A strong lifetime hierarchy is observed in the case of charm hadrons:
$\tau(D^+)\simeq 2.3~\tau(D_s)\simeq 2.5~\tau(D^0)\simeq 5~\tau(\Lambda_c^+)$.
This hierarchy is predicted to scale with the inverse
of the heavy quark mass squared and
the $B$ hadron lifetimes are expected to differ by only
10-20\%\cite{Bigi,Neubert}.

  The technique used by SLD takes advantage of the excellent 3-D
vertexing capabilities of the VXD to reconstruct the decays
inclusively. The goal is to reconstruct and identify all the
tracks originating from the $B$ decay chain. This then allows
charged and neutral $B$ mesons to be separated by simply measuring
the total charge of tracks associated with the $B$ decay.

  The analysis \cite{blrtopol} uses the inclusive topological vertexing
technique described earlier.
In the hadronic $Z^0$ event sample, we select 20,783 $B$ decay candidates
by requiring $M > 2$ GeV and $D > 1$ mm.
The sample is divided into 7942 neutral and 12841 charged vertices
corresponding to reconstructed decays with total charge
$Q = 0$ and $Q = \pm 1, 2, 3$,
respectively, where $Q$ is the charge sum of all tracks associated with
the vertex.
MC studies show that
the ratio between $\bu$ and $\bd$ decays in the charged sample is 1.55 (1.72)
for VXD2 (VXD3), and the ratio between
$\bd$ and $\bu$ decays in the neutral sample is 1.96 (2.24) for
VXD2 (VXD3)\footnote{Reference to a specific state
(e.g., $B^+$) implicitly includes its charge conjugate (i.e., $B^-$).}.

\begin{figure}[htb]
  \centering
  \epsfxsize12cm
  \leavevmode
  \epsfbox{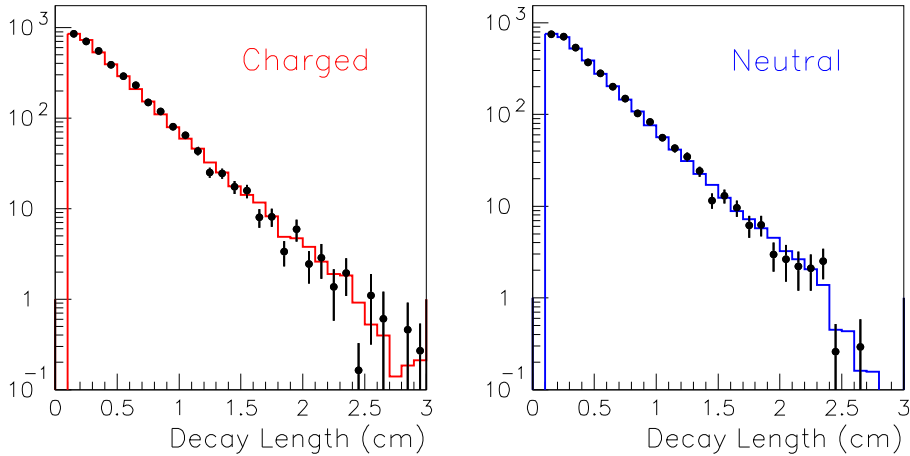}
  \caption{Decay length distributions for data (points) and best fit
           Monte Carlo (histogram).}
  \label{Fig_dkltop}
\end{figure}

  The $\bu$ and $\bd$ lifetimes are extracted with a simultaneous
binned maximum likelihood fit to the decay length distributions of
the charged and neutral samples (see Fig.~\ref{Fig_dkltop}).
The maximum likelihood fit yields lifetimes of
$\tau_{\bu} = 1.698\pm0.040(\mbox{stat})\pm0.046(\mbox{syst})$ ps,
$\tau_{\bd} = 1.581\pm0.043(\mbox{stat})\pm0.061(\mbox{syst})$ ps,
with a lifetime ratio of
$\tau_{\bu}/\tau_{\bd}
    = 1.072^{+0.052}_{-0.049}(\mbox{stat})\pm0.038(\mbox{syst})$.
The main contributions to the systematic error come from uncertainties
in the detector modeling, $\bs$ lifetime, $b$-baryon fraction,
fit systematics, and MC statistics.

These measurements are among the best currently available
and confirm the expectation that the $\bu$ and $\bd$ lifetimes
are nearly equal.

\subsection*{$\bmix$ Mixing}
\label{Sec_bmix}

  Transitions between flavor states $B^0 \leftrightarrow \overline{B^0}$
take place via second order weak interactions ``box diagrams.''
The oscillation frequency $\Delta m_d$ for $\bdmix$ mixing
depends on the CKM matrix element
$\left| V_{td} \right|$ for which little is known experimentally.
Theoretical uncertainties are significantly reduced
for the ratio between $\Delta m_d$ and $\Delta m_s$.
Thus, combining measurements of the oscillation frequency of
both $\bdmix$ and $\bsmix$ mixing translates into a measurement of
the ratio $|V_{td}| / |V_{ts}|$.

  Experimentally, a measurement of the time dependence of $\bmix$
mixing requires three ingredients: (i) the $B$ decay proper time has
to be reconstructed, (ii) the $B$ flavor at production
(initial state $t = 0$) needs to be determined, as well as (iii) the $B$
flavor at decay (final state $t = t_{\rm{decay}}$).
At SLD, the time dependence of $\bdmix$ mixing has been measured using
four different methods. All four use the same initial state tagging
but differ by the method used to either reconstruct the $B$ decay or
tag its final state.

  Initial state tagging takes advantage of the large polarization-dependent
forward-backward asymmetry in $\Zbb$ decays as described above.
For left- (right-) handed electrons and
forward (backward) $B$ decay vertices, the initial quark is tagged as
a $b$ quark; otherwise, it is tagged as a $\overline{b}$ quark.
The initial state tag can be augmented by using a momentum-weighted
track charge (see $A_b$ measurement above) in the hemisphere opposite
that of the reconstructed $B$ vertex.
These two tags are
combined to yield an initial state tag with 100\% efficiency and
effective average right-tag probability of 82\%
(for $\langle P_e \rangle = 77\%$).

  Two $\bdmix$ mixing analyses use topological vertexing
(see $R_b$ measurement) to reconstruct $B$ decays.
The $b/\bar{b}$ flavor tag is performed by using either a kaon-charge tag
(as in the $A_b$ analysis) or
by exploiting the $B \to D$ cascade charge structure.
This latter tag is a novel technique developed by SLD
which relies on a ``vertex charge dipole'' defined as
$\delta q = (\sum^+ w_i L_i) / (\sum^+ w_i)-(\sum^- w_i L_i) / (\sum^- w_i)$,
where the first (second) term is a sum over all positive (negative) tracks
in the vertex and the quantity $L_i$ corresponds to the longitudinal
separation between the IP and the point of closest approach of track $i$
to the vertex line-of-flight. The weight $w_i$ is
inversely proportional to the uncertainty in $L_i$.
Two other analyses select semileptonic decays and thus use the lepton
charge for the $b/\bar{b}$ flavor tag.

The time dependence of $\bdmix$ mixing is measured
from the fraction of decays tagged as mixed as a function of
decay length or proper time.
The four measurements are combined
to produce the following SLD average:
$\Delta m_d = 0.525 \pm 0.043(\mbox{stat}) \pm 0.037(\mbox{syst})$ ps$^{-1}$,
consistent with the world average value of
$0.472 \pm 0.018$ ps$^{-1}$.
Further details about the above measurements may be found
in Ref.~\cite{Willocq} and references therein.

\begin{figure}[htb]
  \centering
  \epsfxsize9cm
  \leavevmode
  \epsfbox{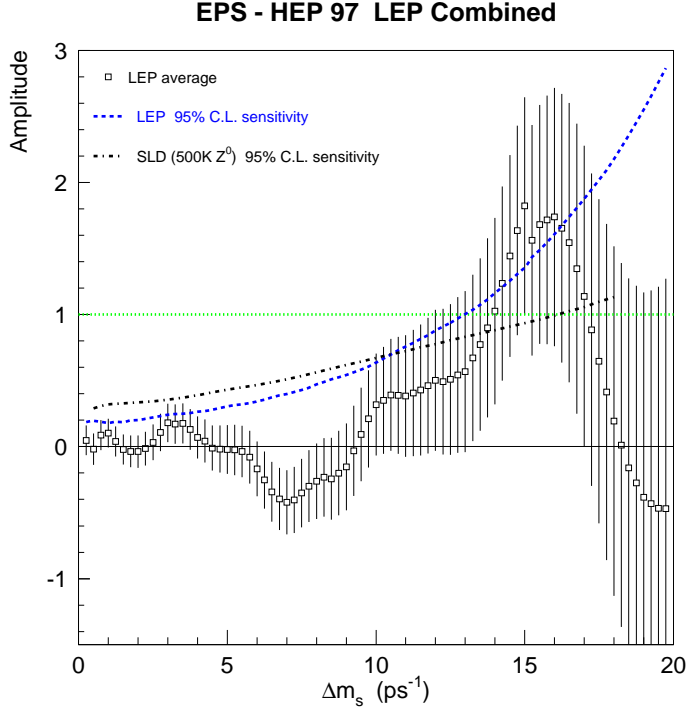}
  \caption{Comparison of sensitivity to $\bsmix$ mixing as a function of
   $\Delta m_s$ for the current LEP average (dashed curve)
   and the projected SLD average for 500K hadronic $Z^0$ (dash-dotted curve).
   Also shown are the current LEP measurements (data points).}
  \label{Fig_bsmix}
\end{figure}

  The above techniques (except for the kaon tag) can be extended to
study $\bsmix$ mixing. Recent studies at LEP indicate that the oscillation
frequency is very large: $\Delta m_s > 10.2$ ps$^{-1}$ at the 95\% C.L.
Therefore, excellent proper time, and thus decay length, resolution
is required to improve that limit or observe oscillations.
With VXD3, SLD expects to achieve decay length resolutions
of 80-100 $\mu$m, i.e. a factor of $\sim 3$ better than those obtained at LEP.
Given a sample of 500K hadronic $Z^0$ decays with VXD3, a limit of
$\Delta m_s > 16$ ps$^{-1}$ can be set at the 95\% C.L.,
see Fig.~\ref{Fig_bsmix}.

\section*{Summary and Prospects}
\label{Sec_sum}

  Using samples of 150K and 50K hadronic $Z^0$ decays
collected in 1993--95 and 1996, the SLD Collaboration has produced precise
and/or unique tests of the Standard Model.
These analyses take advantage of
the large longitudinal electron beam polarization,
the small and stable SLC beam spot,
the high-resolution 3-D pixel vertex detector,
and the particle identification capabilities of the Cherenkov
Ring Imaging Detector.

  Many measurements relying on precise tracking will benefit greatly
from the increased resolution and coverage of the upgraded vertex
detector (VXD3) installed before the 1996 run.
SLD expects to collect another 300K to 400K $Z^0$
by the end of the 1997-98 run.
With that sample, SLD will surpass the precision achieved by the combined LEP 
measurements of $\sinsqth$ and $A_b$, and approach the same precision for
$R_b$.
Furthermore, the superior resolution will enable SLD to significantly
increase the sensitivity to $\bsmix$ mixing beyond that currently
attained by the LEP experiments.

\end{document}